\documentclass[conference]{IEEEtran}
\usepackage{graphicx}
\usepackage[utf8]{inputenc}
\ifCLASSINFOpdf
\else
\fi
\graphicspath{ {Images/} }
\usepackage{listings}

\usepackage{hyperref} 
\hypersetup{
     hidelinks
}

\begin{document}
\title{UEFI BIOS Accessibility for the Visually Impaired}

\author{\IEEEauthorblockN{Rafael R. Machado}
\IEEEauthorblockA{Department of Computing at Sorocaba (DComp)\\
Federal University of São Carlos\\
Sorocaba, São Paulo, Brazil\\
Email: rafaelrodrigues.machado@gmail.com}
\and
\IEEEauthorblockN{Gustavo M. D. Vieira}
\IEEEauthorblockA{Department of Computing at Sorocaba (DComp)\\
Federal University of São Carlos\\
Sorocaba, São Paulo, Brazil\\
Email: gdvieira@ufscar.br}
}


%Telephone: (800) 555--1212, Fax: (888) 555--1212}


\maketitle

\begin{abstract}
%\boldmath
    People with some kind of disability face a high level of difficulty for everyday tasks because, in many cases, accessibility was not considered necessary when the task or process was designed. An example of this scenario is a computer's BIOS configuration screens, which do not consider the specific needs, such as screen readers,  of visually impaired people. This paper proposes the idea that it is possible to make the pre-operating system environment accessible to visually impaired people. We report our work-in-progress in creating a screen reader prototype, accessing audio cards compatible with the High Definition Audio specification in systems running UEFI compliant firmware.
\end{abstract}

\begin{IEEEkeywords}
Accessibility, BIOS, Extensible Firmware Interface, Firmware, High Definition Audio, Screen Reader, UEFI.
\end{IEEEkeywords}

% \fi
\IEEEpeerreviewmaketitle

\section{Introduction}

% Paragraph: Accessibility in the current world
The word inclusion has been widely used to denote the idea that all people are equal and, regardless of any special need, should have the same opportunities.
In order to guarantee these rights to persons with disabilities, the United Nations Organization has created the \textit{Convention on the Rights of Persons with Disabilities}~\cite{hendricks2007convention}. This convention aims to make humanity stop seeing people with disabilities as incapable and worthy of charity, starting to treat them as members capable of having an active participation in society with rights and duties, as can be found in the foreword, items \textit{e)} and \textit{n)} of the convention~\cite{hendricks2007convention}:

\begin{quote}
e) Recognizing that disability is an evolving concept and that disability results from the interaction between people with disabilities and barriers to attitudes and the environment that prevent their full and effective participation in society on an equal basis with other people.
\end{quote}

\begin{quote}
n) Recognizing the importance for people with disabilities of their individual autonomy and independence, including the freedom to make their own choices.
\end{quote}

This idea is strongly linked to the concept of accessibility, which aims to create the necessary ways for people, regardless of having some special need or not, to achieve a given task with the highest possible quality. In this context, the concept of \emph{universal usability} has been created, with the objective of enabling more than 90\% of people to use a certain technology without great difficulties \cite{meiselwitz2009universal, shneiderman2000universal}. 
Examples of universal accessibility are accessible staircases, which provide easier access for wheelchair users but also can make the life of other users easier and more productive. For instance, accessible staircases can help someone pushing a baby stroller or an elderly person who has difficulties when climbing stairs. In short, everyone benefits when universal accessibility is taken into account.
There are several places were the concept of universal usability  can be applied, and one of them is software. With an increasingly connected world, it's important to remember that people with all levels of education, knowledge and impairments are becoming users of  software and  deserve the attention of the software creators.

There is a growing number of information technology professionals who have some type of visual impairment. Among these professionals we can find programmers, network experts and a large number of professionals working with computer maintenance, among other specialties. 
In order to allow people with some type of visual impairment to perform computer related tasks, \emph{screen readers} have been developed, which are software that aim to help the user navigate a visual interface through sound instructions~\cite{allen1981voice}. For example, a user that is using a screen with three buttons could navigate between them through a pre-defined command (often the tab key), and with the help of the screen reader, listen the name of each button that is currently selected. One of the first software of this type was AccessDos in the 1990s, which was used to make the DOS operating system accessible \cite{vanderheiden2008ubiquitous}.

Unfortunately, not all computer related tasks have tools that allow accessibility to all people. One example of vital software that completely lacks any accessibility support is the pre-operating system configuration screens (popularly known as BIOS). This is due to the fact that the pre-OS environment lacks a driver for the audio card, so it is not possible for a screen reader to run. Considering the previously mentioned fact that many visually impaired professionals work with computer maintenance, this can negatively impact the ability of a skilled but visually impaired user to perform simple tasks such as the configuration of a RAID card on a server, the setting of the real time clock or selecting the OS to be loaded on a dual boot system.
These are examples of a myriad of tasks that can be done quickly in a pre-OS environment, without the need for an OS on the system. However, these tasks can be performed only with the help of another person who is not visually impaired, thus not allowing total autonomy of the visual impaired user.

Research indicate that there are around 285 million people with some type of visual impairment in the world, of whom 39 million are blind~\cite{VisualImpairmentWorld}, so the lack of accessibility in the pre-OS applications represents a clear discrimination against these people, because they are not free to use their computers in the same way as people who do not have these conditions.
The companies responsible for the development of firmwares seem not to be interested in creating this kind of accessibility. After all, only a small part of the population is visually impaired, and of this set very few people are computer literate at a level that using a screen reader would make some sense in pre-OS environments. Unfortunately, companies do not realize that when an application becomes accessible, all of the extra expense required for its development is payed by a larger number of users interested on their product~\cite{sherman2001cost, W3CStudyAccessibilityCostBack}.

This paper proposes the idea that it is possible to make the pre-OS environment accessible to visually impaired people. Our long term goal is the development of a screen reader prototype for pre-OS environments and here we report our work-in-progress toward that goal.
In Section~\ref{overview} we outline the architecture of the proposed propotype.
The first step of the construction of this prototype is the development of an audio driver for pre-OS environment, thus allowing sound applications to be developed in firmware.
Given the complexities between the various manufacturers of pre-OS environments and hardware, this work was developed with focus on the firmware that follows the Unified Extensible Firmware Interface (UEFI) specification~\cite{UefiSpec}. In Section~\ref{uefi} we describe this specification in more detail and how we intend to enhance it with the creation of a suitable sound driver. For the same reasons of the selection of UEFI as a base standard, we decided to start with audio cards that follow the High Definition Audio (HDA) specification~\cite{HighDefinitionSpec}. In Section~\ref{hda} we describe the HDA specification and our efforts in creating a driver for this specification. 
We conclude in Section~\ref{conclusion} by listing the missing steps to fully realize the creation of the proposed prototype.

\subsection{Related Work}
\label{related}

Nowadays there are several tools focused on accessibility for the visually impaired. Among them, screen readers are used when the user has very little or no sight at all~\cite{allen1981voice}.
The function of this type of assistive application is to guide the visually impaired user by describing the visual interface he is operating, as well as all actions performed by him.
This allows the users to accomplish all possible tasks within the environment, such as reading documents, accessing e-mails and any other type of activity. This includes advanced tasks such as software development and system configuration changes and maintenance.

\begin{figure}[htb]
\centering
\includegraphics[width=7.01cm, height=9.92cm]{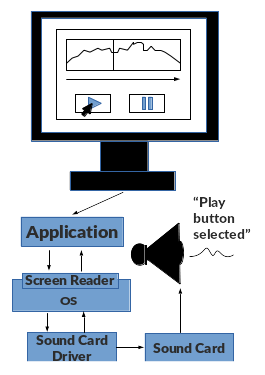}
\caption{Screen reader architecture}
\label{fig:ScreenReaderArq}
\end{figure}

To create a screen reader, several pieces of software must work together, such as the sound drivers, the graphical interface frameworks, the voice synthesizer, etc.~\cite{bidarra2013development}. Fig.~\ref{fig:ScreenReaderArq} shows how some of these components interact, mostly mediated by the operating systems.
Screen readers are available for many environments, encompassing most OSs in use today, as shown in Table~\ref{ScreenReadersAvailableToday}.

%Imagem Screen Reader

\begin{table}[htb]
%\fontsize{11}{11}\selectfont
\centering
\caption{Screen readers}
\label{ScreenReadersAvailableToday}
\begin{tabular}{|l|l|}
\hline
\textbf{Operating System} & \textbf{Screen Reader}                                                           \\ \hline
Windows                   & \begin{tabular}[c]{@{}l@{}}JAWS \cite{JAWS}\\ myReader \cite{myReader}\\ Microsoft Narrator \cite{Narrator}\\ NVDA \cite{NVDA}\end{tabular} \\ \hline
Linux                     & \begin{tabular}[c]{@{}l@{}}SUSE-Blinux \cite{Blinux}\\ Orca \cite{Orca}\end{tabular}                       \\ \hline
Mac OS                    & \begin{tabular}[c]{@{}l@{}}Proloquo2Go \cite{Proloquo2Go}\\ Voice Over \cite{VoiceOver}\end{tabular}                    \\ \hline
Android                   & Talk Back \cite{TalkBack}                                                                        \\ \hline
\end{tabular}
\end{table}

These projects have the same target and, besides some usability changes, all provide the possibility for the user to have an understanding of the environment they are using through audio instructions. They also have options to speed up the software's speak cadence, so the user does not expend too much time when listening the instructions. For a non-trained listener, a high-speed setup would never be understood, proving that the visual impaired users compensate any kind of deficiency with an increase on their listening capabilities.

From a technical point of view, all these screen readers have in common the fact that they rely on the services provided by an operating system. These services range from hooks to the graphical interface frameworks to the sound infrastructure, including sound card drivers, among many others.
Unfortunately, there is no screen reader able to be executed at the pre-OS environment because of the limited services provided. This is somewhat surprising considering the sophistication of the current generation of UEFI firmware, where the user interacts with the system through a  mouse-driven graphical user interface.

\section{Towards an Accessible BIOS}
\label{overview}

\begin{figure}[htb]
\centering
\includegraphics[width=7.01cm, height=9.92cm]{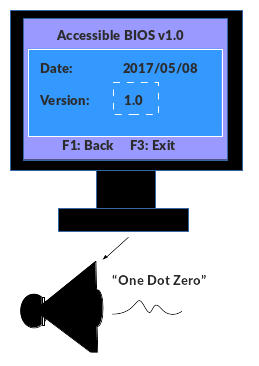}
\caption{Accessible BIOS}
\label{fig:AccessibleBios}
\end{figure}

The addition of a screen reader can be considered a first step of what in future will be an accessible UEFI BIOS, so visual impaired people can use the computers at all levels: operating system and pre-OS environment. From a user's perspective the system should operate as shown in Fig.~\ref{fig:AccessibleBios}. After entering at the system's BIOS, the user would be able to select each field at the screen, and when a given field receives the focus its content would be presented as a sound information, making possible for the user to navigate on the screen following the voice commands.

\begin{figure}[htb]
\centering
\includegraphics[width=7cm, height=8cm]{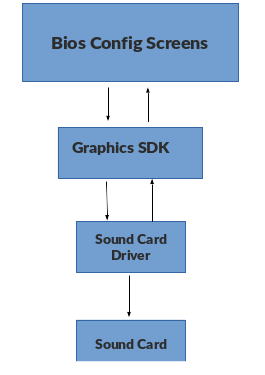}
\caption{Screen Reader Architecture}
\label{fig:Architecture}
\end{figure}

To achieve this objective we propose to integrate in the UEFI environment a sound infrastructure with a basic API tied to the UEFI graphics SDK, as shown in Fig.~\ref{fig:Architecture}. With this foundation in place we can hook the screen elements of the SDK with sound output operations and create a screen reader tailored to the pre-SO environment.

In the remainder of this section we describe the current state of this research, focusing on the creation of a sound output infrastructure.

\subsection{Unified Extensible Firmware Interface (UEFI)}
\label{uefi}

Since the 1960s, the firmware responsible for initializing the devices in a computer was written entirely in Assembly language.
Over time, the equipment began to become increasingly complex, with the inclusion of new types of controllers and technologies.  These firmware, which were no longer trivial software, started to become a tangle of lines of code in Assembly.
To try to minimize this complexity a group of companies created the EFI (Extensible Firmware Interface) concept that later, after some changes, become the UEFI (Unified Extensible Firmware Interface)~\cite{UefiSpec}. This standard defines  the interfaces a firmware should export on a pre-OS environment and allows the creation of portable applications that run in this environment.

\begin{figure}
\centering
\includegraphics[width=250px]{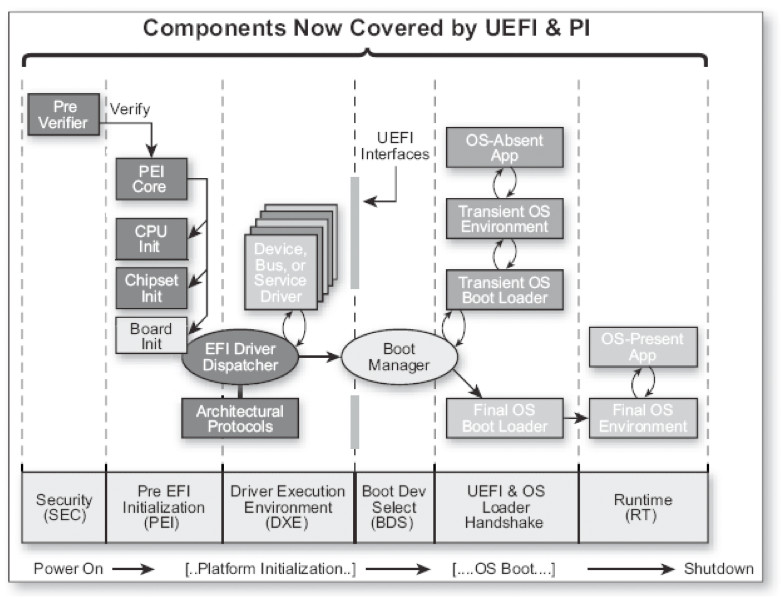}
\caption{PI/UEFI boot process \cite{UefiBootProcess}}
\label{fig:UefiInit}
\end{figure}

In addition to the UEFI specification, another very important specification for the evolution of computer firmware is the PI (Platform Initialization) specification, which defines the steps to be followed during the boot process, and the modules to be loaded at each step on equipment whose firmware is compliant with the specifications.
The steps defined by the PI/UEFI specification are shown in Fig.~\ref{fig:UefiInit}.
Each of the steps presented has a very specific function in the boot process, making this concept very modular and expandable.  Table~\ref{table:AssetStatusBoot} shows the function of each step in the device's boot process.

\begin{table}[!htb]
\centering
\caption{Boot Phases}
\label{table:AssetStatusBoot}
\begin{tabular}{|l|l|}
\hline
Step & Assignments \\ \hline
SEC & \begin{tabular} [c] {@ {} l @ {}} - Handle all system boot events \\ - Create a temporary location to be used as memory \\ - Ensure security requirements for the next step (root of trust) \\ - Pass important information to the next step \end{tabular} \\ \hline
PEI & \begin{tabular} [c] {@ {} l @ {}} - Initiate permanent memory \\ - Perform memory division for specific function (memory map) \\ - Other firmware may be stored (option roms) \\ - Pass the control to the next step \end{tabular} \\ 
\hline
DXE & \begin{tabular} [c] {@ {} l @ {}} - Produce Boot Services \\ - Produce Runtime Services \\ - Discover and run the DXE drivers in the correct order \end{tabular} \\ 
\hline
BDS & \begin{tabular} [c] {@ {} l @ {}} - Implement platform boot policy (which device will start first) \end{tabular} \\ 
\hline
\end{tabular}
\end{table}

Note that the compatibility with the PI/UEFI specification is not restricted to notebooks and desktops. Any equipment may have firmware that follows this specification. One can find mobile phones, tablets and development kits that have firmware compatible with the specification, presenting only some necessary changes given cost and hardware restrictions. With this, portability between equipment and platforms becomes a much simpler task than it was used to be.

The DXE step is the time at which the drivers will be loaded during the initialization of a given system with firmware compliant with the PI specification, thus enabling components and peripherals to be started correctly. For example, in this step SATA controllers, PCI/PCI-e devices, video cards, among other numerous types of controllers and existing peripherals are initialized. In this work we propose to create a sound card driver for the UEFI environment, that will be loaded in the DXE phase. This driver will export a sound interface to all other applications running in the pre-OS environment. 

Since most of the systems currently available  already have an integrated sound card attached to the PCI-e bus, this driver would take advantage of the PCI-e protocols defined by the UEFI specification, making the complexity of such a driver much smaller. To further simplify the construction of the driver and increase its portability, we target HDA audio specification as we describe in the next section. We have created a proof of concept application to help in testing the driver being developed, since in the UEFI environment there is no difference between an application and a driver with respect to hardware access. The conversion of this conceptual application to a DXE driver would be straightforward.

\subsection{High Definition Audio (HDA)}
\label{hda}

Among various sound card architectures, a very popular one in the Intel High Definition Architecture (HDA). This specification defines a standard of communication to be followed by manufacturers who are interested in being compatible with existing Intel platforms. Among other things, this specification defines a communication interface to be adopted by the sound card manufacturers as well as the forms of interaction between the platform and the audio equipment. The specification includes the definition of the controller's register set, the physical description of the interconnections, codec programming model, and architecture components of the codecs~\cite{HighDefinitionSpec}.

This architecture defines two types of devices, controllers and codecs. These two types of hardware work together in the system, enabling the playback of the desired audio. 
The best definition of these concepts come from the HDA specification~\cite{HighDefinitionSpec}:

\begin{LaTeXdescription}
\item[Controller:] The High Definition Audio controller is a bus mastering I/O peripheral, which is
attached to system memory via PCI or other typical PC peripheral attachment host interface. It
contains one or more DMA engines, each of which can be set up to transfer a single audio ``stream''
to memory from the codec or from memory to the codec depending on the DMA type. The
controller implements all the memory mapped registers that comprise the programming interface.

\item[Codec:]  One or more codecs connect to the link. A codec extracts one or more audio streams from
the time multiplexed link protocol and converts them to an output stream through one or more
converters.
\end{LaTeXdescription}

The \emph{verbs} are the commands to be send to the codec, so the codec can be controlled and configured. The verbs are defined by the HDA specification, so a system with a HDA compatible audio needs to have a codec able to answer to all these commands. Examples of these commands are GetParameter (0xF00), Amplifier Gain Mute (0xB--, 0x3--) and Beep Generation Control (0xF0A, 0x70A). Most of the verbs have two ids, one to set the information and another one to retrieve it. All verbs can be found at page 216 of the HDA specification~\cite{HighDefinitionSpec}.

As concrete example,  we have used for development a ASUSPRO P2540UA laptop~\cite{AsusPro}.
This system has a High Definition Audio 1.0a compatible codec, that provides all the sound functionality of the machine. In particular, the codec used to test the prototype was a CX20752 Low-Power High Definition Audio CODEC, from Conexant Inc.~\cite{ConexantCodec}.

\subsection{Prototype Application}
\label{poc}

This first hurdle to the development of sound capabilities in the UEFI environment is that  the UEFI specification has no provisions for a sound protocol. This way we had to start the development in the lower layers. In particular, we had to start with the PCI Express bus, that have a interface defined in  page 638 of the UEFI specification~\cite{UefiSpec}, as shown in Fig.~\ref{fig:pci}.

\begin{figure}
\begin{verbatim}
GUID
#define 
EFI_PCI_ROOT_BRIDGE_IO_PROTOCOL_GUID \
{0x2F707EBB,0x4A1A,0x11d4,0x9A,0x38,\
0x00,0x90,0x27,0x3F,0xC1,0x4D}
\end{verbatim}
\caption{PCI-e interface}
\label{fig:pci}
\end{figure}

Using this low level interface we had first to establish a way to communicate with the HDA controller. To do that, a parser of the PCI configuration space of the chipset was created. This parser was developed using the PCI protocols defined by the UEFI specification and all the bytes were parsed following the chipset specification~\cite{IntelControllerHubSpec}. The HDA controllers are always located at PCI 0:27:0 (this means bus 0, device 27, function 0).  Fig.~\ref{fig:wcr32} shows sample code used to write 32 bit values to the controller registers.

\begin{figure}
\begin{verbatim}
EFI_STATUS WriteControllerRegister32 (
  PCI_HDA_REGION* PcieDeviceConfigSpace,
  UINT64 Offset, UINT32 Value) {

UINTN handleCount = 0;
EFI_STATUS Status = EFI_SUCCESS;
EFI_PCI_ROOT_BRIDGE_IO_PROTOCOL* 
  rootBridgeProtocol = NULL;
EFI_HANDLE* detectedHandles = NULL;

//fix base addres, from device 00 27 00 
//(hda controller)
UINT64 HdaControllerBar = 
  (PcieDeviceConfigSpace->HDBARL 
                          & 0xFFFFFFF0);

Status = 
  gBS->LocateHandleBuffer(ByProtocol,
    &gEfiPciRootBridgeIoProtocolGuid,
    NULL, &handleCount, 
    &detectedHandles);

if(!EFI_ERROR(Status)) {

 Status = 
   gBS->OpenProtocol(detectedHandles[0],
    &gEfiPciRootBridgeIoProtocolGuid,
    (VOID**) &rootBridgeProtocol,
    gImageHandle, 
    NULL,
    EFI_OPEN_PROTOCOL_BY_HANDLE_PROTOCOL);

 if(!EFI_ERROR(Status)) {
 
   Status = 
       rootBridgeProtocol->Mem.Write(
         rootBridgeProtocol,
         EfiPciWidthUint32,
         HdaControllerBar + Offset,
         1,
         (VOID*) &Value);

   gBS->CloseProtocol(detectedHandles[0],
     &gEfiPciRootBridgeIoProtocolGuid,
     gImageHandle,
     NULL);
  }
 }
return Status;
}
\end{verbatim}
\caption{WriteControllerRegister32}
\label{fig:wcr32}
\end{figure}

Once the communication with the HDA controller is set up, we were able to access the codec itself. To send commands to the codec the HDA specification defines some registers that are not mandatory, but that were present and operational on the system used for development. These registers are named Immediate Command Output Interface (ICOI), Immediate Response Input Interface (ICII) and Immediate Command Status (ICIS).
To send a command to the codec the software must write the verb on the ICOI register, and check the ICIS register to detect when the command was processed. The result of the command will be placed in the ICII register.  To execute these commands, the function  WriteControllerRegister32 (Fig.~\ref{fig:wcr32}) is executed receiving as parameter the offset of the ICOI, ICII and ICIS registers, according to the PCI-e configuration space of the HDA controller.

% Removi essas linhas pois a sua definição acima me parece suficiente. Estou mantendo pois você pode querer usar depois.
%
%The formal definition of these registers from the HDA Specification is:
%
%\emph{"The Immediate Command Output and Immediate Command Input registers are optional %registers which provide a Programmed I/O (PIO) interface for sending verbs and receiving %responses from codecs. These registers can be implemented in platforms not suited for DMA %command operations. If implemented, these registers must not be used at the same time as the %CORB and RIRB command/response mechanisms, as the operations will conflict."} %\cite{HighDefinitionSpec}

To validade correct communication with the codec we used two simple verbs: GetParameter (0xF00) and Beep Generator Control (0xF0A, 0x70A).
The GetParameter verb is used to get the details of each widget present in the codec. We were able to validate the results after sending this verb to specific nodes in the codec comparing the result with the chipset datasheet.  
For example, the node 0 at the codec should return the information contained in Table~\ref{table:GetParameter} when the GetParameter verb is send to it, and we were able to observe exactly these values at the pre-OS environment with the prototype application.

\begin{table}[!htb]
\centering
\caption{Node 0 Responses}
\label{table:GetParameter}
\begin{tabular}{|l|l|l|l|l|}
\hline
\textbf{Description} & \textbf{Verb ID} & \textbf{Parameter} & \textbf{Response} &  \textbf{Comments} \\ \hline
Vendor ID            & F00h             & 00h                & 14F1510Fh       & CX20452           \\ \hline
Revision ID          & F00h             & 02h                & 0x00100100         & Revision B0.      \\ \hline
\end{tabular}
\end{table}

\begin{figure}[!htb]
\centering
\includegraphics[width=200px, height=140px]{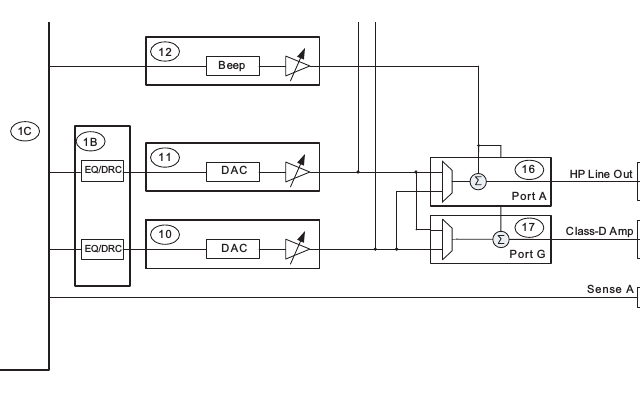}
\caption{Beep Generator}
\label{fig:BeepGenerator}
\end{figure}

The other verb tested used a node in the codec that has a very simple and well defined  behavior. Fig.~\ref{fig:BeepGenerator} shows a block diagram representing the codec's nodes, including one named Beep Generator, at position 0x12, that allows the generation of audible beeps. We have sent the verb Beep Generator Control to this node, along with appropriate  frequencies to be processed, and as a result beeps were heard as expected.
This way we were able to validate the feasibility of accessing the sound card hardware in the pre-OS environment.

\section{Conclusion}
\label{conclusion}

In this paper we proposed the idea that it is possible to make the pre-operating system environment accessible to visually impaired people, and reported our work-in-progress in creating a screen reader prototype for UEFI BIOS with HDA sound cards. We have shown that the sound cards are accessible at the pre-OS environment, and that the development of a screen reader solution for such an environment is possible.

As we reported only a work-in-progress, many questions remain toward our goal of creating a working screen reader prototype. For example, how to handle DMA, and the fact that the PI/UEFI is executed on a single thread. Moreover, work will be needed to convert the developed code in a UEFI DXE driver,  develop  a graphic interface tool kit for the UEFI environment, having accessibility tags to be executed on focus events, and after that a screen reader. 
However, the hardest part would probably be to make the BIOS manufacturers recognize that visual impaired people should be remembered. The technical part, as shown, seems to be totally achievable.

% or

%\appendices
%\blindtext

\appendices

%\section*{Acknowledgment}


\ifCLASSOPTIONcaptionsoff
  \newpage
\fi



\bibliographystyle{IEEEtran}
\bibliography{references}
%\begin{thebibliography}{1}


%\end{thebibliography}

%\begin{biography}[{\includegraphics[width=1in,height=1.25in,clip,keepaspectratio]{mshell}}]{Michael Shell}

%\begin{IEEEbiography}[{\includegraphics[width=1in,height=1.25in,clip,keepaspectratio]{picture}}]%{John Doe}
%\blindtext
%\end{IEEEbiography}


%\vfill


\end{document}